\documentclass{article}
\usepackage{spconf,amsmath,graphicx}
\usepackage{cite}
\usepackage{enumitem}
\usepackage{color}
\usepackage{svg}
\usepackage{booktabs}
\usepackage{amssymb}
\usepackage{multirow}

\usepackage{url}
\usepackage{dsfont} 
\usepackage{adjustbox}
\usepackage{xcolor}
\usepackage[linesnumbered,ruled,vlined]{algorithm2e}
\usepackage{hyperref}
\usepackage[normalem]{ulem}

\usepackage{cleveref} 
\crefrangelabelformat{Sec.}{#3#1#4--#5\crefstripprefix{#1}{#2}#6}

\usepackage{tikz}

\title{Neural Audio Fingerprint for High-specific Audio Retrieval\\based on Contrastive Learning}

\address{\textsuperscript{1}Cochlear.ai, \textsuperscript{2}Seoul National University,\textsuperscript{3}SK Telecom}

\name{\begin{tabular}{c}Sungkyun Chang\textsuperscript{1}, Donmoon Lee\textsuperscript{1,2}, Jeongsoo Park\textsuperscript{1}, Hyungui Lim\textsuperscript{1}, \\
Kyogu Lee\textsuperscript{2}, Karam Ko\textsuperscript{3}, and Yoonchang Han\textsuperscript{1}\end{tabular}}

\begin{document}
\ninept
\maketitle
\begin{abstract}
Most of existing audio fingerprinting systems have limitations to be used for high-specific audio retrieval at scale. In this work, we generate a low-dimensional representation from a short unit segment of audio, and couple this fingerprint with a fast maximum inner-product search. To this end, we present a contrastive learning framework that derives from the segment-level search objective. Each update in training uses a batch consisting of a set of pseudo labels, randomly selected original samples, and their augmented replicas. These replicas can simulate the degrading effects on original audio signals by applying small time offsets and various types of distortions, such as background noise and room/microphone impulse responses. In the segment-level search task, where the conventional audio fingerprinting systems used to fail, our system using 10x smaller storage has shown promising results. Our code and dataset are available at \url{https://mimbres.github.io/neural-audio-fp/}.
\end{abstract}
\begin{keywords}
acoustic fingerprint, self-supervised learning, data augmentation, music information retrieval
\end{keywords}
\section{Introduction}
\label{sec:1}

Audio fingerprinting is a content summarization technique that links short snippets of unlabeled audio contents to the same contents in the database\cite{haitsma2002highly}. The most well-known application is the music fingerprinting system\cite{haitsma2002highly, wang2003industrial, cano2005review, baluja2008waveprint, cotton2010audio, hon2015audio, gfeller2017now} that enables users to identify unknown songs from the microphone or streaming audio input. Other applications include detecting copyrights\cite{cano2005review}, deleting duplicated contents\cite{burges2005using}, monitoring broadcasts\cite{allamanche2001audioid, haitsma2002highly}, and tracking  advertisements\cite{jiang2019audio}.

General requirements for audio fingerprinting system are \textit{discriminability} over a huge number of other fingerprints, \textit{robustness} against various types of acoustic distortions, and \textit{computational efficiency} for processing large-scale database. To achieve these requirements, most of conventional approaches\cite{haitsma2002highly, wang2003industrial, cano2005review, baluja2008waveprint, cotton2010audio, six2014panako, hon2015audio} employed a novelty function to extract sparse representations of spectro-temporal features from a pre-defined audio window. These sparse representations, or acoustic landmarks\cite{cotton2010audio}, used to be coupled with binary hashing algorithms\cite{haitsma2002highly, wang2003industrial, Gionis99similaritysearch} for scalable search in \textit{hamming} space. 

Still, the representation learning approach to audio fingerprinting has not been discovered well. \textit{Now-playing}\cite{gfeller2017now}  has been pioneering work in the direction. They trained a neural network using semi-hard triplet loss, which derived from face recognition\cite{schroff2015facenet}. In their setup\cite{gfeller2017now}, \textit{Now-playing} could identify songs within 44 h long audio database. In our benchmark, we replicate this semi-hard triplet approach and compare it with our work in a new setup: high-specific audio retrieval in a 180 times larger database.

\begin{figure}
 \centerline{
    \includegraphics[width=\columnwidth]{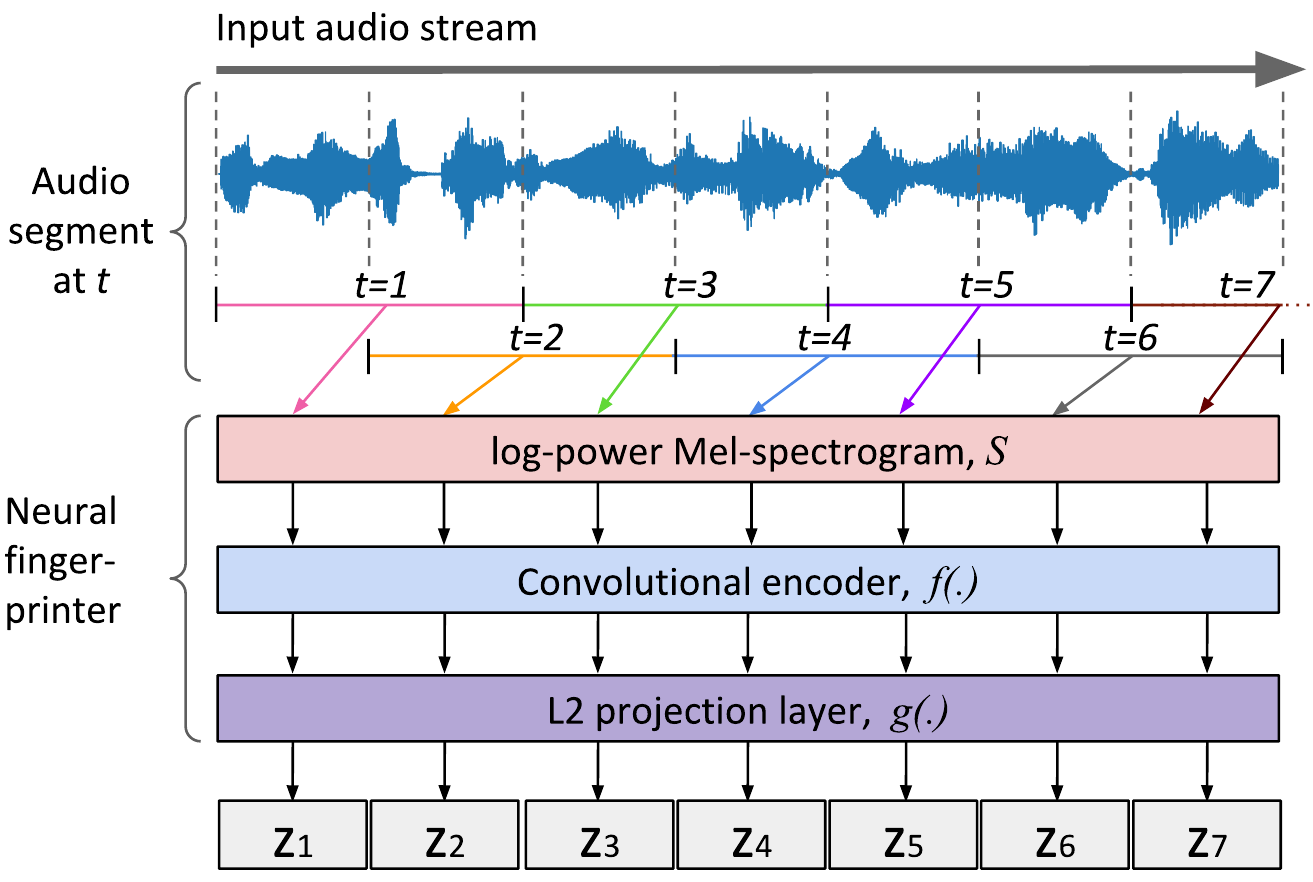}
    }
 \caption{
 Overview of the neural audio fingerprinter. We generate segment-wise embeddings $z_t \in \mathcal{Z}$ that can represent a unit segment of audio from the acoustic features $S$ at time step $t$. In our framework, each segment can be searched by maximum inner-product. 
} 
 \label{fig:fp}
\end{figure}

We present a neural audio fingerprinter for robust high-specific audio retrieval based on contrastive learning.
Our fingerprinting model in Figure~\ref{fig:fp} differs from the prior works in three key aspects:
\begin{itemize}[topsep=1pt,itemsep=2pt,parsep=2pt, leftmargin=*]
    \item Prior works\cite{haitsma2002highly, wang2003industrial, cano2005review, baluja2008waveprint, cotton2010audio, six2014panako, hon2015audio, gfeller2017now} have focused on song-level audio retrieval from a music excerpt; we challenge a high-specific audio search by allowing miss-match less than 250 ms from a few seconds input.
    \item We introduce the contrastive learning framework for simulating maximum inner-product search (MIPS) in mini-batch.
    \item We employ various types of data augmentation methods for generating acoustic distractors and show their benefits to training a robust neural audio fingerprinter.
\end{itemize}

\section{Neural audio fingerprinter}
\label{sec:2}





Our neural audio fingerprinter in Figure~\ref{fig:fp} transforms and maps the segment-level acoustic features into $L^2$-normalized space, where the inner-product can measure similarities between segments. It consists of a pre-processor and neural networks.

As a first step, input audio $\mathcal{X}$ is converted to time-frequency representation  $\mathcal{S}$. It is then fed into convolutional encoder $f(.)$  which is based on the previous study \cite{gfeller2017now}. Finally, $L^2$-normalization is applied to its output through a linear projection layer $g(.)$. 
Thus, we employ $g\circ f:\mathcal{S} \mapsto \mathcal{Z}^d$ as a segment-wise encoder that transforms $\mathcal{S}$ into d-dimensional fingerprint embedding space $\mathcal{Z}^d$.
The d-dimensional output space $\mathcal{Z}^d$ always belongs to \textit{Hilbert} space $L^2(\mathbb{R}^d)$: the cosine similarity of a pair unit such as $\text{cos}(z_a, z_b)$ becomes inner-product $z_a^T z_b$, and due to its simplicity, $L^2$ projection has been widely adopted in metric learning studies\cite{schroff2015facenet, chen2020simple, gfeller2017now}.

The $g \circ f(.)$ described here can be interpreted as a reorganization of the previous audio fingerprinting networks\cite{gfeller2017now} into the common form employed in self-supervised learning (SSL)\cite{oord2018representation, chen2020simple, chen2020big, baevski2020wav2vec}.
However, our approach differs from the typical SSL that throws $g(.)$ away before fine-tuning for the target task: we maintain the self-supervised $g(.)$ up to the final target task. 


\subsection{Contrastive learning framework} 
\label{subsec:objective}
\begin{figure}[t!]
 \centerline{
    \includegraphics[width=\columnwidth]{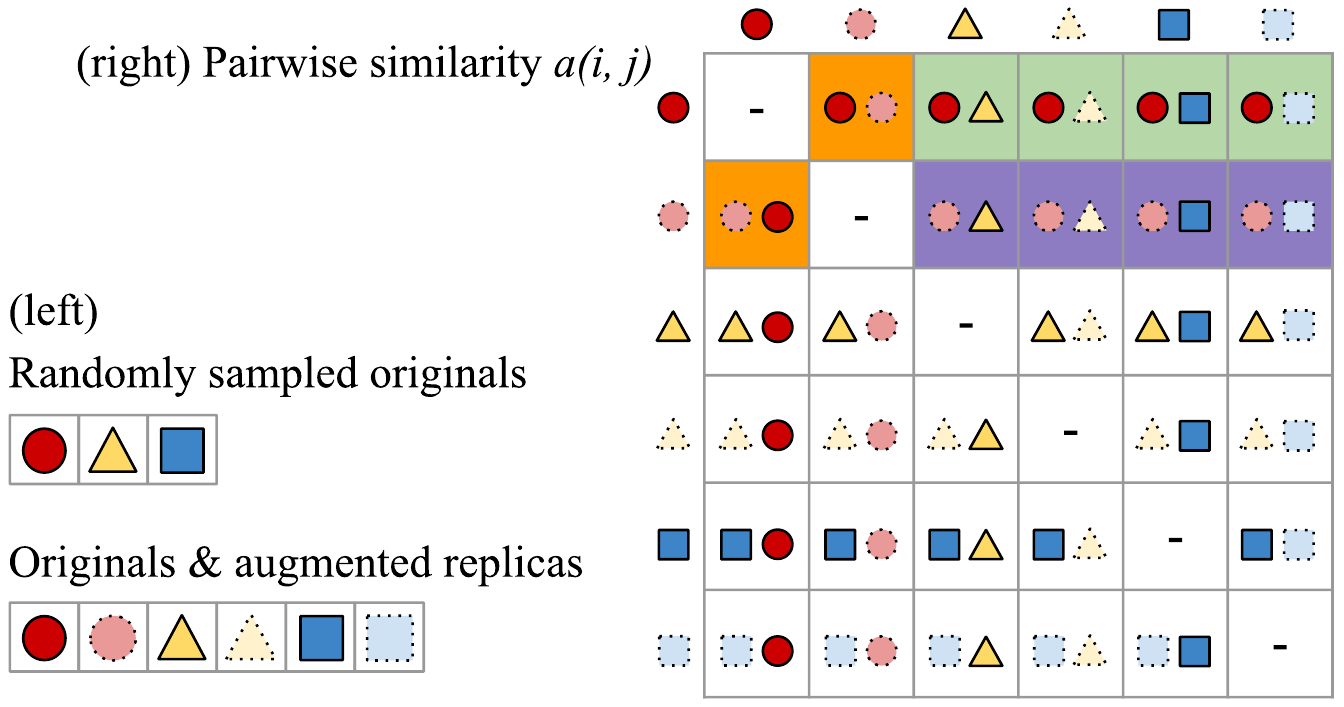}
    }
 \caption{
Illustration of the contrastive prediction task in Section 2.1. (left) Batch size $N=6$. We prepare $N/2$ pairs of original/replica. 
The same shapes with solid/dashed lines represent the positive pair of original/replica, respectively. (right) Each element in the matrix represents pairwise similarity. In each row, a prediction task can be defined as classifying a positive pair (one of the orange squares) against the negative pairs (green or purple squares) in the same row.}
 \label{fig:simCLR_loss}
\end{figure}



As mentioned earlier, we can use the inner-product as a measure of similarity between $z_t \in \mathcal{Z}^d$ for any time step $t$. 
Without losing generality, searching the most similar point (*) of database V = $\{v_i\}$ for a given query $q$ in $\mathcal{Z}^d$ space can be formulated as maximum inner product search (MIPS), $v_i^{\ast} := \text{arg\,max}_i(q^{\top} v_i)$.


We simulate MIPS in a mini-batch setup that takes into account various acoustic distortions and input frame mismatches occurring in the fingerprint task.
A mini-batch with the size of $N$ consists of $N/2$ pairs of  $\{s^{\text{org}}, s^{\text{rep}}\}$. $s^\text{org}$ is the time-frequency representation of sampled audio and $s^\text{rep}$ is the augmented replica of $s^\text{org}$, where $s^\text{rep}=\mathcal{M}_{\alpha}(s^\text{org})$. $\mathcal{M}_{\alpha}$ is an ordered augmentation chain that consists of multiple augmentors with the random parameter set $\alpha$ for each replica.
In this configuration, the indices of original examples are always odd, and that of replicas are even.
Therefore, the batch-wise output of $f\circ g(s)$ can be $ \{z_{2k-1}^{\text{org}}, z_{2k}^{\text{rep}}\}_{k=1}^{2/N}$.

We give each $k$-th example a chance to be an anchor (or a query in MIPS) to be compared with all other examples excluding itself in the batch. We calculate the pairwise inner-product matrix between all elements in the batch $\{z_i\}_{i=1}^{N}$ as  $a(i,j) = z_i^{T} z_j$ for $\forall i,j\in \{1,2,...,N \}$ as Figure~\ref{fig:simCLR_loss}.
Then, we define the contrastive prediction task for a positive pair of examples $(i,j)$ as:
\begin{equation}
\label{eq:NT_xent}
\ell(i,j) = -\text{log} \frac{\text{exp}(a_{i,j} / \tau)}
{\sum_{k=1}^{N} \mathds{1}(k \neq i) \: \text{exp}(a_{i,j} / \tau)}.
\end{equation}
$\mathds{1}(.) \in \{0, 1\}$ is an indicator function that returns $1$ iff $(.)$ is true, and $\tau > 0$ denotes the temperature\cite{hinton2015distilling} parameter for softmax.
We employ Equation~\ref{eq:NT_xent} to replace MIPS from the property: computing the top-$k$ ($k$=1 in our setup) predictions in the softmax function is equivalent to the MIPS. A similar approach is found in \cite{chen2018learning}.
The total loss $\mathcal{L}$ averages $l$ across all positive pairs, both $(i, j)$ and $(j, i)$:  
\begin{equation}
\label{eq:final_loss}
    \mathcal{L} = \frac{1}{N} \sum_{k=1}^{N} [\ell(2k-1, 2k), \ell(2k, 2k-1)].
\end{equation}
Updating rules are summarized in Algorithm \ref{alg:trainig_fp}. 

\newcommand\mycommfont[1]{\footnotesize\ttfamily\textcolor{blue}{#1}}
\SetCommentSty{mycommfont}

\SetKwInput{KwConfig}{Config}
\SetKwInput{KwInput}{Input}
\SetKwInput{KwVariables}{Variables}
\SetKwInput{KwAugmentor}{Augmentor} 
\SetKwInput{KwNets}{Nets}
\SetArgSty{textrm}

\begin{algorithm}[t]
\label{alg:trainig_fp}
    \small
    \DontPrintSemicolon
    \caption{Training of neural audio fingerprinter}
    \KwConfig{even number of batch size $N$, temperature $\tau$}
    \KwVariables{input $s$, representation $z \in \mathbb{R}^d$}
    \KwAugmentor{$\mathcal{M}_{\alpha}(.)$ with parameters $\alpha$}
    \KwNets{encoder $f(.)$, $L^2$ projection layer $g(.)$}
    \For{each sampled mini-batch $\{s_k\}^{N/2}_{k=1}$}    
    {
        \For{$\forall k\in\{1,...,N/2\}$}
        {
            $z^{\text{org}}_k = g\circ f(s_k)$\\ 
            $z^{\text{rep}}_k = g\circ f(\mathcal{M}({s}_k))$
            
        }
        
        $z = \{z^{\text{org}}_1, z^{\text{rep}}_1,...,z^{\text{org}}_{N/2},z^{\text{rep}}_{N/2} \}$
        
        \For{$\forall i\in\{1,...,N\}$ and $\forall j\in\{1,...,N\}$}
        {
            $a_{i,j}=z_i^{\top} z_j$  \tcc*{Pairwise similarity}
            $\ell(i,j) = \text{NTxent}(a_{i,j}, \tau)$ \tcc*{Eq.\eqref{eq:NT_xent}}
        }
        Update $f,g$ to minimize $\mathcal{L} \approx \frac{1}{N}\displaystyle\sum_{i=1}^{N}{\ell}$ \tcc*{Eq.\eqref{eq:final_loss}}
    }
    \Return{fingerprinter $g\circ f(.)$}
\end{algorithm}
It is worth comparing our approach to SimCLR\cite{chen2020simple} for visual representation. Our approach differs from SimCLR on how to construct positive pairs. We use \textit{\{original, replica\}}, whereas SimCLR uses \textit{\{replica, replica\}} from the same original source. In our case, the anchor is already given because the database will always store the clean source, so it can be more important to learn the consistent relation between the original and its replica over all other negatives.

\subsection{Sequence search}
\label{subsec:search}

Our model trained by simulating MIPS is optimized for segment-level search. 
In the case of searching for a query sequence $\{Q_{i=0}^L\}$ consisting of $L$ consecutive segments:
We first gather the top $k$ segment-level search results indices $I_{q_{i}}$ for each $q_i$ from the DB.
The offset is then compensated by $I'_{q_i} = I_{q_i} - i$.
The set of candidate indices $c \in C$ is determined by taking unique elements of $I'_{q_i}$.
The sequence-level similarity score is the sum of all segment-level similarities from the segment index range $[c, c+L]$, and the index with the highest score is the output of the system.




\section{Experimental setup}
\label{sec:5}

\subsection{Dataset}

The main experiment in Table~\ref{tab:seg_result} is reproduceable with the following three data sets, which are isolated from each other.
\begin{itemize}[topsep=1pt,itemsep=2pt,parsep=2pt, leftmargin=*]
\item Train (10K-30s): A subset of the \texttt{fma\_medium}\cite{fma_dataset} consisting of 30 s audio clips from a total of 10K songs. 
\item Test-Dummy-DB (100K-full-db): a subset of the \texttt{fma\_full}\cite{fma_dataset} consisting of about 278 s audio clips from a total of 100K songs. We scale the search experiment with this.
\item Test-Query/DB (500-30s): Test-DB is another subset of the \texttt{fma\_medium}, which is 500 audio clips of 30 s each. Test-Query was synthesized using Test-DB as directed in Section~\ref{subsec:eval_protocol}.
\end{itemize}



\subsection{Data pipeline with augmentation chain}
\label{ssec:augmentation}

A batch consists of $\{x^{\text{org}}, x^{\text{rep}}\}$ pairs.
Each $x^{\text{rep}}$ is generated from its corresponding $x^{\text{org}}$ through augmentation steps as following order:
\begin{itemize}[topsep=1pt,itemsep=2pt,parsep=2pt, leftmargin=*]
    \item Time offset modulation: To simulate possible discrepancies in real world search scenarios, we define positive examples as 1 s audio clips with an offset of up to $\pm$200 ms. We first sample 1.2 s of audio and then $\{x^{\text{org}}, x^{\text{rep}}\}$ are chosen by random start positions.
    \item Background mixing: A randomly selected noise in the SNR range of [0, 10] dB is added to the audio to reflect the actual noise. The noise dataset consists of 4.3 h of a subset of AudioSet\cite{gemmeke2017audio} and 2.3 h of pub and cafe noise recorded by us. The AudioSet was crawled within \textit{subway}, \textit{metro}, and \textit{underground} tags with no music-related tags. Each dataset is split into 8:2 for train/test.
    \item IR filters: To simulate the effect of diverse spatial and microphone environments, microphone and room impulse response (IR) are sequentially applied by convolution operation. Public microphone\cite{micir} and spacial\cite{jeub2009binaural} IR dataset are split into 8:2 for train/test.
    \item Cutout\cite{devries2017improved} and Spec-augment\cite{park2019specaugment} are applied after extracting log-power Mel-spectrogram features, such that $\{s^{\text{org}}, s^{\text{rep}}\}$. Unlike other augmentations, we uniformly apply a batch-wise random mask to all examples in a batch including $s^{\text{org}}$. The size and position of each rectangle/vertical/horizontal mask is random in the range [1/10, 1/2] the length of each time/frequency axis.  
\end{itemize}

\subsection{Network structure}
\label{subsec:net_structure}
\newcommand{\F}[1]{\textsf{#1}}
\newcommand{\FF}[1]{#1}
\newcommand{\<}{{\vartriangleleft}}

\begin{table}[t!]
  \renewcommand{\arraystretch}{1.4}
  \footnotesize
  \caption{Fingerprinter (FP) network structure in Section \ref{subsec:net_structure}.} 
  \label{fig:architecture}
  \centering
  \begin{tabular}{l}
    \toprule
    $\F{SC}^{o\gets i}_{k*s}(.)$
    $:= \F{ReLU} \< \F{LN} \< \F{C}^{o\gets i}_{k'*s'} \< \F{ReLU} \< \F{LN} \< \F{C}^{o\gets i}_{k*s}(.)$ \\
    \midrule
    $\FF{f}(.)$
    $:=$
    $\F{SC}^{h\gets h}_{3*2} {\vartriangleleft} \F{SC}^{h\gets 4d}_{3*2}{\vartriangleleft}\F{SC}^{4d\gets 4d}_{3*2}{\vartriangleleft}\F{SC}^{4d\gets 2d}_{3*2}{\vartriangleleft}$\\
    $\;\;\;\;\;\;\:\: \F{SC}^{2d\gets 2d}_{3*2}$ ${\vartriangleleft} \F{SC}^{2d\gets d}_{3*2}{\vartriangleleft} \F{SC}^{d\gets d}_{3*2} {\vartriangleleft}\F{SC}^{d\gets 1}_{3*2}(.)$\\
    \midrule
    $\FF{g}(.)$
    $:= \F{L2} \< \F{Concat} \< \F{C}^{1\gets u}_{1*1} \< \F{ELU} \< \F{C}^{u\gets v}_{1*1} \< \F{Split}^{h\//d}(.)$\\
    \midrule
    $\F{FP}$
    $:= \FF{g} \< \FF{f}(\F{input}:=s_t)$ \\
    \bottomrule
  \end{tabular}
\end{table}
In Table~\ref{fig:architecture}, a space-saving notation $\F{C}^{o\gets i}_{k*s}$ denotes Conv2d with input channel $i$, output channel $o$, kernel size $1\times$$k$, and stride $1\times$$s$. The $k'$ and $s'$ denote rotation as $k\times$$1$ and $s\times$$1$.
$\F{Split}^{h\//d}$ splits input dimension $h$ into $d$ parts of each output dimension $v=h/d$. $g\<f(.)$ is $g(f(.))$. The network parameters $\{d,h,u,v\}$ are in Table~\ref{tab:exp_setup}.
\begin{itemize}[topsep=1pt,itemsep=2pt,parsep=2pt, leftmargin=*]

    \item \textbf{Convolutional encoder $\boldsymbol{f(.)}$}: 
    $f(.)$ takes as input a log-power Mel-spectrogram $s_t$ with a time step $t$ represnting 1s audio captured by 50\% overlapping window.
    $f(.)$ consists of several blocks containing spatially separable convolution (SC)\cite{mamalet2012simplifying} followed by a layer normalization (LN)\cite{ba2016layer} and a ReLU activation.
    \item \textbf{$\boldsymbol{L^2}$ projection layer $\boldsymbol{g(.)}$}: We take the split-head from the input  embeddings and pass it through the separate Linear-ELU-Linear layers for each split as in previous studies\cite{gfeller2017now, lai2015simultaneous}. After concatenating the multi-head outputs, we apply $L^2$-normalization. 
    

    
\end{itemize}

\subsection{Implementation details}
\begin{table}[t!]
  \footnotesize 
  \caption{Shared configurations for experiments}
  \label{tab:exp_setup}
  \centering
  \begin{tabular}{ll}
    \toprule
    Parameter& Value \\
    \midrule
    Sampling rate & 8,000 Hz\\
    STFT window function & \textit{Hann}\\
    STFT window length and hop & 1024, 256\\
    STFT spectrogram size $F\times T$ & $512\times T (T=32)$\\
    log-power Mel-spectrogram size $F\prime\times T$ & $256\times T (T=32)$\\
    Dynamic range & 80 dB \\
    Frequency \{min, max\} & \{300, 4,000\} Hz\\
    Fingerprint \{window length, hop\} & $\{1s, 0.5s\}$ or $\{2s, 1s\}$\\
    Fingerprint dimension $d$ &  64 or 128 \\
    Network parameters $\{h,u,v\}$ & $\{1024,32,h/d\}$\\
    Batch size $N$ & 120 or 320 or 640\\
  \bottomrule
\end{tabular}
\end{table}
The replication of \textit{Now-playing} and our work shared the short-time Fourier transform (STFT) settings listed in in Table~\ref{tab:exp_setup}. Note that, due to ambiguity in the previous study\cite{gfeller2017now}, the STFT parameters were set by us. We trained \textit{Now-playing} using online semi-hard triplet loss\cite{schroff2015facenet} with the margin $m=0.4$ and batch size $N=320$.    

We trained our model using LAMB\cite{lamb_optimizer} optimizer, which performed 2 pp better than Adam\cite{adam_optimizer}   with the 3 s query sequence for batch size $N \geq 320$. In practice, Adam worked better only for $N\leq240$. The learning rate had an initial value of $\text{1e-4}\cdot N/640$ with cosine decay without warmup\cite{goyal2017accurate} or restarts\cite{loshchilov2016sgdr}, then it reached a minimum value of 1-e7 in 100 epochs.
The temperature in Eq.\ref{eq:NT_xent} was $\tau = 0.05$, and we did not observe a meaningful performance change in the range $[0.01, 0.1]$.
The training finished in about 30 h with a single \textit{NVIDIA RTX 6000} GPU or \textit{v3-8} Cloud TPUs.

The search algorithm in Section~\ref{subsec:search} was implemented using an open library\cite{johnson2019billion}. We used the inverted file (IVF) index structure with product quantizer (PQ) as a non-exhaustive MIPS. The IVF-PQ had 200 centroids with the code size of $2^6$, and 8-bits per index. In this setting, the loss of recall remained below 0.1\% compared to the exhaustive search of 100K songs ($\approx$56M segments) database. 


\subsection{Evaluation protocol}
\label{subsec:eval_protocol}
\begin{itemize}[topsep=1pt,itemsep=2pt,parsep=2pt, leftmargin=*]
\item Evaluation metric: To measure the performance in segment/song-level search in Section~\ref{sec:4}, we use \textit{Top-1 hit rate(\%)}:
\begin{equation}
\footnotesize
     100\times \frac{\textit{(n of hits @Top-1)}}{(\textit{n of hits @Top-1) + (n of miss @Top-1})},
\end{equation}
which is equivalent to \textit{recall}.
In Table~\ref{tab:seg_result}, \textit{exact match} is the case when the system finds the correct index in database. We further define the tolerance range for \textit{near match} as $\pm$500 ms.
\item Test-Query generation: 2K query-sources for each \{1, 2, 3, 5, 6, 10\} s length are randomly cropped from Test-DB containing 500 clips of 30s each. 
Each query is synthesized through the random augmentation pipeline as described in Section~\ref{ssec:augmentation}. Note that we exclude Cutout and Spec-augment. The default SNR range is [0, 10] dB. 
We make sure that the data used for background mixing and IR as unseen to our model by isolating them from training set.
\end{itemize}

\section{Results and Discussion}
\label{sec:4}

\begin{table}[t]
\caption{
Top-1 hit rate (\%) of large-scale (total of 100K songs) segment-level search. \textit{d} denotes the dimension of fingerprint embedding. \textit{exact match} means that our system finds the exact index. \textit{near match} means a mismatch within $\pm$ 1 index or $\pm$ 500 ms.} 
\label{tab:seg_result}
\centering
\renewcommand{\arraystretch}{1.25}
{\scriptsize %
\begin{tabular}{cclllllll}
\hline
\multirow{2}{*}{Method} & \multirow{2}{*}{$d$}   & \multicolumn{1}{c}{\multirow{2}{*}{\begin{tabular}[c]{@{}c@{}} {\scriptsize match}\end{tabular}}} & \multicolumn{6}{c}{\scriptsize Query length in seconds}\\
\cline{4-9}                                                                                               &                       & \multicolumn{1}{c}{} & 1 s    & 2 s       & 3 s      & 5 s      & 6 s     & \multicolumn{1}{c}{10$s$ }\\
\hline
\multirow{2}{*}{\begin{tabular}[c]{@{}c@{}} \textit{Now-playing}\\ (replicated)\end{tabular}} & \multirow{2}{*}{128}  & {\scriptsize exact}  & -  & 44.3     & 60.1    & 73.6    & 81.0   & 86.1\\
                                                                                              &                       & {\scriptsize near}   & -  & 46.8     & 63.5    & 75.2    & 81.6   & 86.3\\
\hline
\multirow{4}{*}{\begin{tabular}[c]{@{}c@{}} \textit{Now-playing} \\ (modified\\ for 1 s unit)\end{tabular}} & \multirow{2}{*}{64}  & {\scriptsize exact} & 25.8      & 58.5      & 69.3      & 78.5      & 81.4      & 87.7\\
                                                                                     &                      & {\scriptsize near}   & 30.9      & 61.3      & 71.2      & 79.5      & 82.2       & 88.3\\ \cline{2-9} 
                                                                                     & \multirow{2}{*}{128} & {\scriptsize exact}  & 26.3     & 58.2      & 69.5      & 78.4      & 81.4       & 87.8\\
                                                                                     &                      & {\scriptsize near}   & 30.9      & 61.1      & 71.8      & 79.8      & 83.0       & 89.2\\
\hline
\multirow{4}{*}{\begin{tabular}[c]{@{}c@{}}This work\\($N$=640)\end{tabular}}    & \multirow{2}{*}{64}  & {\scriptsize exact}   & 54.6      & 78.9      & 85.4      & 90.4      & 92.0       & 94.9\\
                                                                               &                      & {\scriptsize near}    & 61.3      & 81.7      & 86.7      & 90.9      & 92.7       & 95.1\\
\cline{2-9}                                                                    & \multirow{2}{*}{128} & {\scriptsize exact}   & \textbf{62.2} & \textbf{83.2} & \textbf{87.4} & \textbf{92.0} & \textbf{93.3} & \textbf{95.6}\\
                                                                               &                      & {\scriptsize near}    & 68.3      & 84.9      & 88.7      & 92.7      & 94.1       & 95.8\\
\hline
\multirow{2}{*}{\begin{tabular}[c]{@{}c@{}}This work\\($N$=320)\end{tabular}}    & \multirow{2}{*}{128}  & {\scriptsize exact}  & 61.0      & 82.2      & 87.1      & 91.8      & 93.1       & 95.2\\
                                                                                &                      & {\scriptsize near}   & 67.1      & 84.1      & 88.1      & 92.5      & 93.9       & 95.5\\
\hline
\multirow{2}{*}{\begin{tabular}[c]{@{}c@{}}This work\\($N$=120)\end{tabular}}    & \multirow{2}{*}{128}  & {\scriptsize exact}  & 55.9      & 78.8      & 84.9      & 90.9      & 92.2       & 95.3\\
                                                                                &                      & {\scriptsize near}   & 62.3      & 80.9      & 86.3      & 91.5      & 92.8       & 95.5\\
\hline
\multirow{2}{*}{\begin{tabular}[c]{@{}c@{}}This work\\(no aug.)\end{tabular}}    & \multirow{2}{*}{128}  & {\scriptsize exact}  & 0.0      & 0.0      & 0.0      & 0.0      & 0.0       & 0.0\\
                                                                                &                      & {\scriptsize near}   & 0.0      & 0.0      & 0.0      & 0.0      & 0.0       & 0.0\\
\hline
\end{tabular}
} %
\end{table}
\subsection{Experimental results}

The main results are listed in Table \ref{tab:seg_result}. 
Using the same augmentation method,  \textit{Now-playing}\cite{gfeller2017now} based on semi-hard triplet  \cite{schroff2015facenet} took 2 s as a unit audio segment. The modified \textit{Now-playing} with 1 s unit audio segment could be more fairly compared with our works.

\smallskip\noindent
\textbf{VS. \textit{Now-playing} (semi-hard triplet)} Modified \textit{Now-playing} consistently performed better than the replicated \textit{Now-playing}. 
While cutting the dimension in half, this trend was maintained.
Considering that the DB size was the same when the number of fingerprint dimensions was half, it could be seen that constructing DB with 1 s was more advantageous to segment search. The proposed model with a 128-dimensional fingerprint using batch size of 640 always showed the best performance (highlighted in Table~\ref{tab:seg_result}) for any query length. This confirmed that the proposed contrastive learning approach outperformed over the semi-hard triplet approach.

\begin{table}[t!]
\caption{Effect of fingerprint dimension $d$ in 1 s segment search.}
\label{tab:effect_of_dim}
\centering
\renewcommand{\arraystretch}{1.25}
{\scriptsize
\begin{tabular}{p{0.35\columnwidth}p{0.1\columnwidth}p{0.1\columnwidth}p{0.1\columnwidth}p{0.1\columnwidth}}
\hline
Embedding dimension                   & \hfil$d$=16 & \hfil$d$=32 & \hfil$d$=64 & \hfil$d$=128 \\
\hline
Top-1 hit rate@1 s (\%)  & \hfil11.6 & \hfil40.2 & \hfil54.6 & \hfil62.2 \\   
\hline
\end{tabular}
}
\end{table}
\smallskip\noindent
\textbf{Embedding dimension} In Table~\ref{tab:exp_setup}, increasing the embedding dimension $d$: 64$\rightarrow$128 for the modified \textit{Now-playing} did not affect the results significantly.
In contrast, increasing the embedding dimension $d$: 64$\rightarrow$128 for our best model gave us a larger improvement of exact match performance as $\uparrow$7.6 (54.6$\rightarrow$62.2\%) pp for the 1 s query.
This reaffirmed the training benefits of our contrastive learning over the semi-hard triplet, fairly compared  using the same network structure.
In Table~\ref{tab:effect_of_dim}, we further investigated the effect of reducing $d$ to our model with 1 s query length. We could observed a rapid drop in exact match performance while decreasing $d$: 64$\rightarrow$32$\rightarrow$16.

\smallskip\noindent
\textbf{Performance of sequence search} The longer the query sequence, the better the performance in all experiments. In Table~\ref{tab:seg_result}, segment-level hit rate of our best model (highlighted) was increasing as 62.2$\rightarrow$83.4$\rightarrow$92.0$\rightarrow$95.6\% while increasing the query length by almost double. Thus, the longer query length was useful. In Table~\ref{tab:seg_result}, the performance difference between near and exact match result of our best model at 1 s query was 6.1 (62.2 and 68.3\%) pp. This interval decreased immediately as the query length became larger than 1. These results showed that our sequence search method introduced in Section~\ref{subsec:search} was quite effective.

\smallskip\noindent
\textbf{Effect of batch size} The larger the batch size, the better the performance in all experiments. 
In Table~\ref{tab:seg_result}, reducing the batch size $N$: 640$\rightarrow$120 from our best model degraded the exact match performance by $\downarrow$6.3 (62.2$\rightarrow$55.9\%) pp at 1 s query length.
Recent works\cite{chen2020simple, chen2020big, baevski2020wav2vec} on contrastive learning has been consistently reporting similar trends. Our result implicated that the diversity of negative examples existing by large batch could play an important role in the contrastive learning framework.




\smallskip\noindent
\textbf{VS. \textit{Dejavu}}
We compared our work with the open-source project \texttt{Dejavu}\cite{dejavu} based on the conventional method\cite{haitsma2002highly, cotton2010audio} in the song-level search task of smaller (10K-30s) scale.
69.6\% of Top-1 hit rate was achieved with \texttt{Dejavu}, a song-level search engine using a 6 s query. Our best model achieved 99.5\% for song-level hit rate, and exact/near match was 98.9/99.1\% at the 6 s query, respectively. Our model also achieved \{83.6, 95.4, 97.4\}\% exact match at \{1,2,3\} s query. The capacity of fingerprints from \texttt{Dejavu} was about 400 MB, while ours (quantized with 1/4 compression rate) was less than 40 MB for $d$=64. These results suggest that our method has advantages over conventional methods in both performance and scalability.


\subsection{Size of training set, search time and scalability}





The models in Table~\ref{tab:seg_result} were trained with about 70 h dataset. This size was less than 1\% of the total 8K h DB for test. We assumed that using the entire DB for training would be impractical--a huge number of new songs are produced every day.
In additional experiment, we used 10\% of the Test-dummy-DB for training a $d$=64 model. It achieved \{\textit{58.3, 81.1, 86.5, 92.4, 93.4, 96.0}\}\% of Top-1 hit rate for the query sequence of \{\textit{1, 2, 3, 5, 6 ,10}\} s. This  improved $\uparrow$3.7 (54.6$\rightarrow$58.3\%) pp at the 1 s query over the best model with $d$=64 in Table~\ref{tab:seg_result}, still lower than the result of $d$=128. Thus, both $d$ and the amount of training data were the factors affecting performance.

In our best model with $d$=128, the final DB size was about 5.8 GB for 56M segments from total of 100K songs. We report about 1.5 s search time with \textit{i9-10980XE} CPU (in-memory-search), and 0.02 s with GPU for parallel search of 19 segments (= 10 s query). In case of using CPUs, we could observe on-disk-search using the latest SSD with CPU was only twice as slow as in-memory-search. We reserve the industry-level scalability issues for future work.






\subsection{Transfer to down-stream task}
We further investigated the generality of the learned embeddings by performing a downstream task, as in the typical SSL\cite{oord2018representation, chen2020simple, chen2020big, baevski2020wav2vec} settings. By fixing $f(.)$ and fine-tuning a linear classifier, we tried audio genre classification in \textit{GTZAN} dataset with stratified 10-fold cross-validation.
Fine-tuning on the pre-trained embeddings for fingerprint achieved 59.2\% accuracy, while training from scratch achieved only 32.0\%.
This showed that the features encoded by $f(.)$ were linearly interpretable, consistent with other SSL reports\cite{oord2018representation, chen2020simple, chen2020big, baevski2020wav2vec}.
However, our result was slightly lower than the baseline of 61.0\% accuracy using MFCCs+GMM\cite{tzanetakis2002musical}. This might be due to the limitation of the lightweight networks with the relatively short-time analysis window.


\section{Conclusions and future work}
\label{sec:7}
This study presented a neural audio fingerprinter for high-specific audio retrieval. Our model was trained to maximize the inner-product between positive pairs of fingerprints through a contrastive prediction task.
To this end, we explicitly sampled positive pairs to have original--replica relations by applying various augmentations to clean signals.
We evaluated our model in the segment-level search task with a public database of 100K songs. In the experiment, our model performed better than the model with triplet embeddings.  
It was also shown that our work, using 10 times less memory than an existing work, outperformed in song-level search task.
So far, these results have implied that the audio fingerprinting task would inherently have self-supervised learning potentials.
The future direction of this study is to test neural audio fingerprints in industry-scale database and queries from a variety of user devices.





\vfill\pagebreak
\section{Acknowledgement}
We would like to thank the TensorFlow Research Cloud (TFRC) program that gave us access to Google Cloud TPUs.
\bibliographystyle{IEEEbib}
\bibliography{References}

\end{document}